\documentclass[prl,twocolumn,aps,amsmath,nofootinbib,superscriptaddress]{revtex4-1}
\usepackage{graphicx}
\usepackage{bm}
\usepackage{epsfig}
\usepackage{hyperref}

\newif\ifpdf
\ifx\pdfoutput\undefined
\pdffalse 
\else
\pdfoutput=1 
\pdftrue
\fi

\def\OMIT#1{}

\newcommand{\nn}{\nonumber}

\newcommand{\bea}{\begin{eqnarray}}
\newcommand{\eea}{\end{eqnarray}}

\newcommand{\beq}{\begin{equation}}
\newcommand{\eeq}{\end{equation}}

\begin{document}
\ifpdf
\DeclareGraphicsExtensions{.pdf, .jpg}
\else
\DeclareGraphicsExtensions{.eps, .jpg}
\fi

\title{The Casimir Torque on a Cylindrical Gear}

\author{Varun Vaidya}
\affiliation{Department of Physics, Carnegie Mellon University\footnote{Permanent Address},
    Pittsburgh, PA 15213}


\begin{abstract}
We utilize Effective Field Theory(EFT) techniques to calculate the casimir torque on a cylindrical gear in the presence of a polarizable but neutral object. We present results for the energy and torque as a function of angle for a gear with multiple cogs, as well as for the case of a concentric cylindrical gear.
\end{abstract}

\maketitle

\section{Introduction}
The Casimir Force has been a subject of many research papers since the force between two polarizable atoms \cite{CP} was calcuated by Casimir and Polder in 1948. Since then there has been tremendous interest in this field involving the physical effects of this force in various geometrical configurations \cite{KM},\cite{Enk}. With the recent advancement in nanotechnology, there has been an interest in non-contact gears \cite{NC} to determine the torque between two corrugated concentric cylindrical surfaces. As a first step towards this goal, this paper uses the world line effective field theory approach \cite{IW} to calcuate the interaction energy between a cylindrical gear and a polarizable object. This technique is then extended to the case of concentric cylindrical gears with dielectric cogs. The approach mirrors the work done in the context of membranes \cite{SM}, \cite{IP}.
\section{Casimir Torque}

We first consider the simple case of a infinitely long perfectly conducting cylinder with a single dielectric cog which we denote as A(fig.\ref{gear1}). The cylinder is centered at the origin and we have a small polarizable object(B) at a distance r from the origin. Both A and B are neutral and isotropic. For simplicity, we define the z coordinates of A and B to be the same. The relevant scales in the problem are: The size of the cogs (R), the energy gap of the cogs $(\Delta E)$ and the distance between the cogs $r$. We will assume that $(1/\Delta E, R) \ll r$ so that we will be performing an expansion in  $\lambda_1 \equiv R/r$ as well as $\lambda_2 \equiv 1/(\Delta E r)$. The expansion in $\lambda_1$ corresponds to the multipole expansion, while $\lambda_2$ controls corrections arising from exciting the internal degrees of freedom of the cog. We will be working at leading order in both these expansion parameters, though the corrections can be easily accounted for within the effective field theory formalism. Futhermore, we will consider the limit  $\lambda_1\gg \lambda_2$, so that the dominant corrections will come from the multipole expansion, though this is just a formality since we are working at leading order.



In the effective theory formalism \cite{IW} one begins by integrating out the higher scales  $(1/R, \Delta E$), generating a series of higher dimensional operators whose coefficients can be determined by matching. In this way, the cogs are treated as point particles ($A,B$). These particles are taken to be static so that their world lines have no dynamical action. 
 The finite size effects are encoded in higher dimensional operators which reside on the world line and are constructed by writing down the lowest dimensional operators consistent with the relevant symmetries: Lorentz, gauge, and world line reparametarization invariance. At leading order in $\lambda_1$ we have two opeartors \cite{GLR}, so that the action is given by: 

\begin{figure}
\centering
\includegraphics[width=3in]{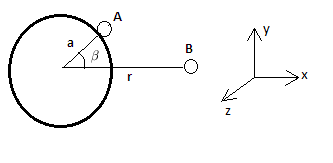}
\caption{Conducting cylindrical gear}
\label{gear1}
\end{figure}

\beq
S_{int}=\int{d\tau\sum_{i=1}^{2}(C_{b_i}\sqrt{v^{2}}F_{\mu\nu}F^{\mu\nu} + \frac{Ce_i}{\sqrt{v^2}}v^\mu F_{\mu\nu}v_{\alpha}F^{\alpha \nu})}.
\eeq

All the information about the internal structure of the cogs is absorbed in the couplings $C_{e_i}$ and $C_{b_i}$ which are determined via a matching procedure. Dimensional analysis tells us that $C_{e_i}$ and $C_{b_i}$ scale as $R^{3}$. The above theory can be made more accurate by adding higher dimensional operators involving more derivatives but each such operator will be suppressed by higher powers of $\lambda_1$. 
By working at the level of the action, one can calculate in arbitrarily complicated geometries as long as the expansion in $\lambda_1$ is well behaved. 
 On the other hand, matching can be done in any simple physical process where the exact result, using the full microscopic theory, can be either calculated, if internal dynamics are understood, or measured otherwise. Matching tells us that the effective couplings $C_{e_i}$ and $C_{b_i}$ are related to the the electric and magnetic polarizabilities ($\alpha_{e_i}/2$) and ($\alpha_{b_i}/2$) respectively \cite{IW}. We consider only electric polarization, since both A and B are stationary such that

\begin{equation}
S_{int}=\int{d\tau\sum_{i=1}^{2} (\frac{-\alpha_{e_i}}{2}E^{2})}.
\end{equation}

The contribution of the magnetic polarizability is suppressed by the velocity  of the internal constituents of the composite particles A and B. 
 We employ the path integral approach to calculate the component of the Casimir interaction energy which contributes to the torque, via the relation 

\beq
<0|e^{-iVT}|0> = \int{DA e^{i(S_{0}+S_{int})}}\nn\\
\eeq

Here $S_{0}$ is the action for the electromagnetic field in the presence of the cylinder without the perturbations, A and B.
V is the total energy which includes both, the self energies of the cogs A and B, and their interaction energy. Out of these two contributions, only the interaction energy of A and B which depends on angle $\beta$ [see Fig.1], contributes to the torque. The leading order contribution to the interaction energy is given by

\bea
&&V_{int}=\frac{-i}{2T}<0|S_{int}^{2}|0>\nn\\
&&=\frac{-i}{4T}(\alpha_{e_1}\alpha_{e_2})\int{d\tau_1\int{d\tau_2<E^{2}(\vec{r'},\tau_1)E^{2}(\vec{r},\tau_2)>}}
\eea

\begin{figure}
\centering
\includegraphics[width=2in]{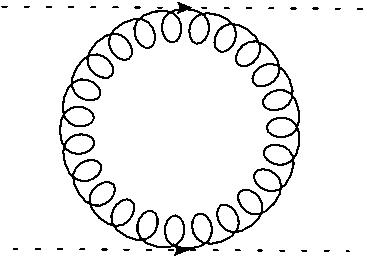}
\caption{Feynman diagram for the relevant term $V_{int}$}
\label{concgear}
\end{figure}

To calculate the energy we need the time ordered two point function(propagator) for each combination of components of the electric field in the presence of a conducting cylinder. We use the fact that the propagator is the Green's function for the equation of motion of the field, calculated in the Feynman prescription. This prescription is later used to perform a wick rotation for evaluating the contour integral for the green's function. Since we have a vector field, we then need to evaluate the greens dyadic (a 3x3 matrix) with the boundary conditions for the electromagnetic field at the surface of the conducting cylinder. This has already been calculated in  \cite{KM} and is given by

\beq
i<E_{i'}(\vec{r'},\tau_1)E_j(\vec{r},\tau_2)>=\int{\frac{d\omega}{2\pi}e^{-i\omega(\tau_1-\tau_2)}(\hat{i'}.\stackrel{\leftrightarrow}{\Gamma}.\hat{j})}
\eeq

where $i,j=\{r,\phi,z\}$.

\bea
\stackrel{\leftrightarrow}{\Gamma}(\omega,\vec{r},\vec{r}')=\sum_{m=-\infty}^{\infty}\int{\frac{dk_z}{2\pi}}
[{\bf MM'^{*}} F_m(r,r')\\\nn  
+ \frac{1}{\omega^{2}}{\bf NN'^{*}}G_m(r,r')]         \chi_{mk_z}(\phi,z)\chi^{*}_{mk_z}(\phi',z')\nn
\eea

where the primed operators act on the primed co-ordinates.

\bea
\chi_{mk_z}(\phi,z)=\frac{1}{\sqrt{2\pi}}e^{im\phi}e^{ik_{z}z}\nn
\\{\bf M}=\hat{r}\frac{im}{r}-\hat{\phi}\frac{\partial}{\partial r}\nn
\\{\bf N}=\hat{r}ik_z\frac{\partial}{\partial r}-\hat{\phi}\frac{mk_z}{r}-\hat{z}d_m\nn
\\d_m=\frac{1}{r}\frac{\partial}{\partial r}r\frac{\partial}{\partial r}-\frac{m^{2}}{r^{2}}\nn
\eea

For $r>r'$,

\bea
F_m(r,r')&=&\frac{\omega^{2}i\pi}{2k_\rho^{2}}H_m(k_{\rho}r)[J_m(k_\rho r')-\frac{J_m'(k_{\rho}a)}{H_{m}'(k_{\rho}a)}H_{m}(k_{\rho}r')]\nn\\
&&-\frac{1}{2|m|k_{\rho}^2}[(\frac{r'}{r})^{|m|}+\frac{a^{2|m|}}{r^{|m|}r'^{|m|}}]\nn
\eea

\bea
G_m(r,r')&=&\frac{\omega^{2}i\pi}{2k_\rho^{2}}H_m(k_{\rho}r)[J_m(k_\rho r')-\frac{J_m(k_{\rho}a)}{H_{m}(k_{\rho}a)}H_{m}(k_{\rho}r')]\nn\\
&&- \frac{1}{2|m|k_{\rho}^2}[(\frac{r'}{r})^{|m|}-\frac{a^{2|m|}}{r^{|m|}r'^{|m|}}]\nn
\eea

\bea
k_{\rho}=\sqrt{\omega^2 - k_{z}^2}\nn
\eea

This is basically an expansion of the Green's function in vector cylindrical harmonics. Another equation that will aid in simplifying the calculation is the Wronskian for the Greens function.

\begin{equation}
J_{m}(k_{\rho}r)H_{m}'(k_{\rho}r)-J_{m}'(k_{\rho}r)H_{m}(k_{\rho}r)=\frac{2i}{\pi k_{\rho}r}\nn
\end{equation}

The coordinates of A and B are $\vec{r'}\equiv$(a,$\phi'$,z) and $\vec{r}\equiv$(r,$\phi$,z) respectively. Since A is located on the surface of the cylinder, the boundary conditions imply that the components of the Greens dyadic that contribute to the energy are 

\bea
V_{int}&=&\frac{-i}{2T}(\alpha_{e_1}\alpha_{e_2})\int{d\tau_1}\int{d\tau_2(<E_{r'}(\vec{r'},\tau_1)E_r(\vec{r},\tau_2)>^2 }+\nn\\
&<&E_{r'}(\vec{r'},\tau_1)E_\phi(\vec{r},\tau_2)>^2+<E_{r'}(\vec{r'},\tau_1)E_z(\vec{r},\tau_2)>^2)
\eea





Evaluating each of the terms gives a fairly complicated expression for the interaction energy.

\begin{widetext}
\bea
&V&_{int}=i(\frac{\alpha_{e_1}\alpha_{e_2}}{2})\int{\frac{d\omega}{2\pi}}\int{\frac{dk_{z_1}}{2\pi}}\int{\frac{dk_{z_2}}{2\pi}}\sum_{m1=-\infty}^{\infty}\sum_{m_2=-\infty}^{\infty}\frac{e^{im_1(\phi-\phi')}}{2\pi}\frac{e^{im_2(\phi-\phi')}}{2\pi}\\\nn& &
[[\frac{im_1k_{z_1}^2}{k_{\rho_1}^2 ar}\frac{H_{m_1}(k_{\rho1} r)}{H_{m_1}(k_{\rho_1} a)}-\frac{im_1\omega^2}{k_{\rho_1}^2 a^2 }\frac{H_{m_1}'(k_{\rho_1} r)}{H_{m_1}'(k_{\rho_1} a)}][\frac{im_2k_{z_2}^2}{k_{\rho_2}^2 ar}\frac{H_{m_2}(k_{\rho_2} r)}{H_{m_2}(k_{\rho_2} a)}-\frac{im_2\omega^2}{k_{\rho_2}^2 a^2 }\frac{H_{m_2}'(k_{\rho_2} r)}{H_{m_2}'(k_{\rho_2} a)}]\\\nn& &
 + [\frac{k_{z_1}^2}{k_{\rho_1} a}\frac{H_{m_1}'(k_{\rho_1} r)}{H_{m_1}(k_{\rho_1} a)}-\frac{\omega^2m_1^2}{k_{\rho_1}^3 a^2 r}\frac{H_{m_1}(k_{\rho_1} r)}{H_{m_1}'(k_{\rho1} a)}][\frac{k_{z_2}^2}{k_{\rho_2} a}\frac{H_{m_2}'(k_{\rho_2} r)}{H_{m_2}(k_{\rho_2} a)}-\frac{\omega^2m_2^2}{k_{\rho_2}^3 a^2 r}\frac{H_{m_2}(k_{\rho_2} r)}{H_{m_2}'(k_{\rho_2} a)}]]
\eea
\end{widetext}

with 

\begin{eqnarray}
k_{\rho_1}=\sqrt{\omega^2 - k_{z_1}^2}\ k_{\rho_2}=\sqrt{\omega^2 - k_{z_2}^2}.\nn
\end{eqnarray}

The contour integral over $\omega$ is to be done using the Feynman prescription. This can be achieved by first doing a counterclockwise rotation in the complex $\omega$ plane effectively going to Euclidean space, which is essentially a wick rotation.\\
Define $\eta=-i\omega$ \ , \ $\lambda_j=-ik_{\rho_j}$ which gives $\lambda_j^2=\eta^2 + k_{z_j}^2$

\begin{equation}
H_{m}(ix)=\frac{2}{\pi}\frac{1}{i^{m+1}}K_{m}(x)\nn
\end{equation}

Here $K_{m}(x)$ is the modified Bessel function of second kind. Define $\beta=\phi-\phi'$ and $y=r/a$, and rescaling the integration variables, the final expression for the energy is given in terms of two terms with definite parity.

\begin{eqnarray*}
V_{int}&=&-(\frac{\alpha_{e_1}\alpha_{e_2}}{2a^7})\int{\frac{d\eta}{2\pi}}\int{\frac{dk_{z_1}}{2\pi}}\int{\frac{dk_{z_2}}{2\pi}}\\& &
\sum_{m1=-\infty}^{\infty}\sum_{m_2=-\infty}^{\infty}\frac{e^{i\beta(m_1+m_2)}}{4\pi^2}[-V_1(m_1,m_2)+V_2(m_1,m_2)]
\end{eqnarray*}

\begin{widetext}
\begin{eqnarray*}
& &V_1(m_1,m_2)=[\frac{m_1k_{z_1}^2}{\lambda_1^2 y}\frac{K_{m_1}(\lambda_1 y)}{K_{m_1}(\lambda_1 )}+\frac{m_1\eta^2}{\lambda_1^2 }\frac{K_{m_1}'(\lambda_1 y)}{K_{m_1}'(\lambda_1 )}][\frac{m_2k_{z_2}^2}{\lambda_2^2 y}\frac{K_{m_2}(\lambda_2 y)}{K_{m_2}(\lambda_2 )}+\frac{m_2\eta^2}{\lambda_2^2 }\frac{K_{m_2}'(\lambda_2 y)}{K_{m_2}'(\lambda_2 )}]\\
& &V_2(m_1,m_2)=[\frac{k_{z_1}^2}{\lambda_1 }\frac{K_{m_1}'(\lambda_1 y)}{K_{m_1}(\lambda_1 )}+\frac{\eta^2m_1^2}{\lambda_1^3 y}\frac{K_{m_1}(\lambda_1 y)}{K_{m_1}'(\lambda_2 )}][\frac{k_{z_2}^2}{\lambda_2 }\frac{K_{m_2}'(\lambda_2 y)}{K_{m_2}(\lambda_2 )}+\frac{\eta^2m_2^2}{\lambda_2^3 y}\frac{K_{m_2}(\lambda_2 y)}{K_{m_2}'(\lambda_2)}]
\end{eqnarray*}
\end{widetext}

\begin{figure}
\centering
\includegraphics[width=3in]{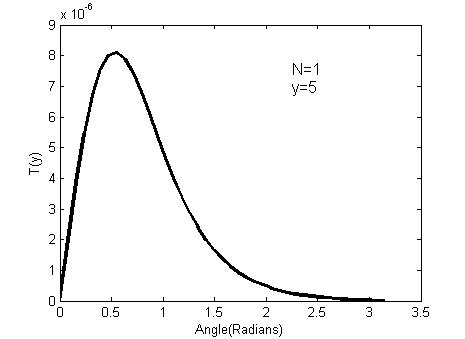}
\caption{T(y,$\beta$) for a gear with a single cog, y=r/a=5}
\label{cg1}
\end{figure}
\begin{figure}
\centering
\includegraphics[width=3in]{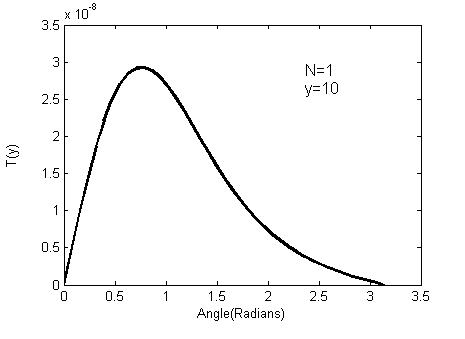}
\caption{T(y,$\beta$) with y=r/a=10}
\label{cg2}
\end{figure}

Using the fact that $K_{m_1}(x)=K_{-m_1}(x)$, it is seen that $V_1$ is odd while $V_2$ is even in $m_1$ and $m_2$. The even term $V_2$ contributes to a an attractive force while the odd terms gives a repulsive one. However, the magnitude of the $V_1$ is much smaller than $V_2$, which still leads to a net attractive force.

\begin{align}
V_{int}=\frac{\alpha_{e_1}\alpha_{e_2}}{4a^7}F(y,\beta)
\end{align}

$F(y,\beta)$ is a dimensionless function of y and $\beta$.
In the numerical evaluation, only the first 6 modes have been included, since the contribution to the energy declines rapidly with increasing mode number. As $y$ gets closer to one we need to include higher modes to maintain the same accuracy.

Similarly, the torque is given by

\begin{align}\label{torque}
Torque=\frac{\alpha_{e_1}\alpha_{e_2}}{4a^7}T(y,\beta)
\end{align} 

We plot $T(y,\beta)$ for $y=5$ and $y=10$ in fig.(\ref{cg1}) and fig.(\ref{cg2}) respectively

\begin{figure}[thb]
\centering
\includegraphics[width=2in]{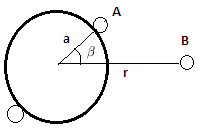}
\caption{Cylinder with two cogs}
\end{figure}

\begin{figure}
\centering
\includegraphics[width=3in]{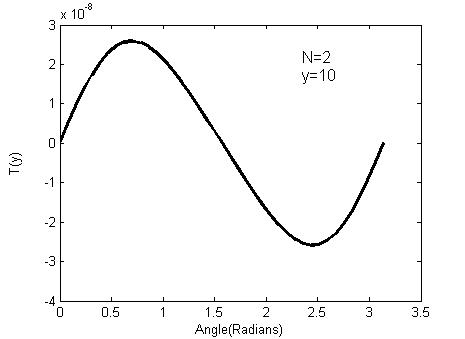}
\caption{T(y,$\beta$) for a two cog gear, y=r/a=10}
\label{cg2t}
\end{figure}

The same logic can be easily extended to the case when the cylinder has multiple cogs, again ignoring the self interaction energy of the cogs. At the same time, the interaction energy of any two cogs on the surface of the cylinder does not contribute to the torque. So the relevant energy is simply the interaction energy of the surface cogs with the off surface one. The corresponding torque is plotted in fig.(\ref{cg2t}).
\section{Concentric gear}

\begin{figure}
\centering
\includegraphics[width=2in]{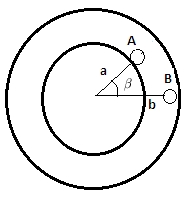}
\caption{concentric cylindrical gear with a single cog}
\label{concgear}
\end{figure}

A similar procedure is followed to obtain the torque in the case of a concentric gear with dielectric cogs. The simplest case is a  gear with one cog as shown in fig.(\ref{concgear}), we have the outer conducting cylindrical shell of radius b. The Green's dyadic is again of the same form with the functions $F_m$ and $G_m$ modified as follows.

\begin{figure}
  \centering
    \includegraphics[width=3in]{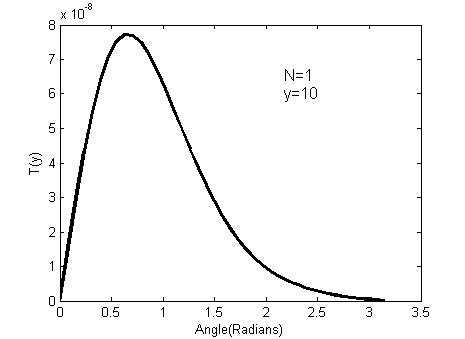}
  \caption{T(y,$\beta$) for a concentric gear with a single cog, y=b/a}
  \label{cc1}
\end{figure}

For $r<r'$,

\bea
F_m(r,r')&=&\frac{\omega^{2}i\pi}{2k_\rho^{2}}[\frac{J_m'(k_{\rho}a)J_m'(k_{\rho}b)H_m'(k_{\rho}a)H_m'(k_{\rho}b)}{J_m'(k_{\rho}b)H_m'(k_{\rho}a)-J_m'(k_{\rho}a)H_m'(k_{\rho}b)}]\nn\\
&& [-\frac{J_m(k_{\rho}r)J_m(k_{\rho}r')}{J_m'(k_{\rho}a)J_m'(k_{\rho}b)} + \frac{J_m(k_{\rho}r)H_m(k_{\rho}r')}{J_m'(k_{\rho}a)H_m'(k_{\rho}b)}\nn\\
&& +\frac{H_m(k_{\rho}r)J_m(k_{\rho}r')}{H_m'(k_{\rho}a)J_m'(k_{\rho}b)} - \frac{H_m(k_{\rho}r)H_m(k_{\rho}r')}{H_m'(k_{\rho}a)H_m'(k_{\rho}b)}]\nn
\eea
\bea
G_m(r,r')&=&\frac{\omega^{2}i\pi}{2k_\rho^{2}}[\frac{J_m(k_{\rho}a)J_m(k_{\rho}b)H_m(k_{\rho}a)H_m(k_{\rho}b)}{J_m(k_{\rho}b)H_m(k_{\rho}a)-J_m(k_{\rho}a)H_m(k_{\rho}b)}]\nn\\
&& [-\frac{J_m(k_{\rho}r)J_m(k_{\rho}r')}{J_m(k_{\rho}a)J_m(k_{\rho}b)} + \frac{J_m(k_{\rho}r)H_m(k_{\rho}r')}{J_m(k_{\rho}a)H_m(k_{\rho}b)}\nn\\
&& +\frac{H_m(k_{\rho}r)J_m(k_{\rho}r')}{H_m(k_{\rho}a)J_m(k_{\rho}b)} - \frac{H_m(k_{\rho}r)H_m(k_{\rho}r')}{H_m(k_{\rho}a)H_m(k_{\rho}b)}]\nn
\eea
\bea
k_{\rho}=\sqrt{\omega^2 - k_{z}^2}\nn
\eea

\begin{figure}
   \centering
       \includegraphics[width=3in]{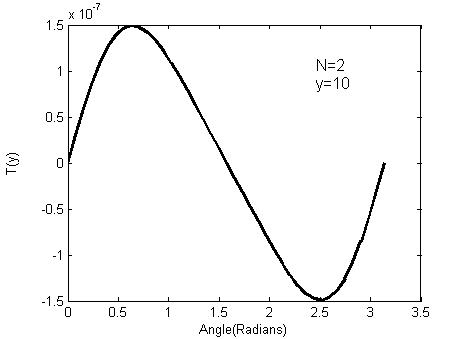}
   \caption{T(y,$\beta$) for a concentric gear with two cogs }
   \label{conc2}
\end{figure}

We have ignored the solutions to the homogeneous equation since they do not contribute to the physical observables. In this case, boundary conditions imply that only the $<E_{r}E_{r'}>$ component of the dyadic will contribute to the torque. The final expression for the torque is again of the same form as eqn.(\ref{torque}), with y now defined as $b/a$. Fig.(\ref{cc1}) shows the torque for the single cog case while fig.(\ref{conc2}) gives the result for the case of two cogs.


\section{Conclusion}
The effective field theory approach is an efficient way of calculating the interaction energy and subsequently the torque. The interaction energy is finite and leads to an attractive force and torque. We have obtained the results for the torque on a cylindrical gear and a concentric one in the regime where the size of the  cogs is much smaller than the distance between them and the energy gap $\Delta E$ of the cogs is much greater than the inverse of the distance between the cogs. The overall normalization of the torque scales linearly in the polarizability of the cogs. Thus the maximum torque will be achieved for cogs with more mobile constituents. However, one must recall that we have assumed that the energy gap is smaller than $1/r$. Should one wish to use this formalism where this condition is not met, then the theory must be augumented to allow for degrees of freedom to live on the cogs as in \cite{abs}. Further work in this area would be of interest. 

\section{Acknowledgements}
This work is supported by DOE contracts DOE-ER-40682-143 and DEACO2-C6H03000. I also thank Ira Rothstein for several discussions on the conceptual aspects of this project.

\end{document}